\documentclass[prl,aps,amssymb,amsmath,reprint,superscriptaddress,showpacs]{revtex4-1}
\usepackage{bm}
\usepackage{graphicx}
\usepackage{color}

\newcommand{\w}{\omega}

\newcommand{\bkd}{{b_k^{\dagger}}}
\newcommand{\bk}{{b_k^{\phantom{\dagger}}}}
\newcommand{\bkpd}{{b_{k'}^{\dagger}}}
\newcommand{\bkp}{{b_{k'}^{\phantom{\dagger}}}}

\begin{document}

\title{Josephson-Kondo screening cloud in circuit quantum electrodynamics} 

\author{Izak Snyman}
\affiliation{Mandelstam Institute for Theoretical Physics, School of Physics, University of the Witwatersrand, Wits, 2050, South Africa}
\author{Serge Florens}
\affiliation{Institut N\'{e}el, CNRS and Universit\'e Grenoble Alpes, F-38042 Grenoble, France}

\begin{abstract}
We show that the non-local polarization response in a multimode circuit-QED setup, devised 
from a Cooper pair box coupled to a long chain of Josephson junctions, provides
an alternative route to access the elusive Kondo screening cloud. 
For moderate circuit impedance, we compute analytically the universal
lineshape for the decay of the charge susceptibility along the circuit, that relates to
spatial entanglement between the qubit and its electromagnetic environment. At large 
circuit impedance, we numerically find further spatial correlations that 
are specific to a true many-body state.
\end{abstract}

\date{\today}

\maketitle

Superconducting circuits constitute at present one of the most 
versatile platforms for quantum engineering, due to their macroscopic
- yet fully coherent - building blocks~\cite{Makhlin}, offering great possibilities in
tunability and design~\cite{Schoelkopf}.
Beyond their promising use in quantum information processing,
such large-scale electrical devices made out of Josephson junctions can 
be viewed as metamaterials~\cite{Zueco,Jung} where light-matter interactions can be explored
in uncharted territory~\cite{Niemczyk,Forn,Astafiev,Abdumalikov,Sanchez}.
These systems allow for unusual physical regimes where the effective fine structure 
constant can become of order one~\cite{LeHur,Goldstein}, so that ultra-strong coupling between a single 
two-level system and a large number of environmental modes leads to 
a vacuum with non-trivial entanglement properties~\cite{Bera1}.
Our aim in this Letter is to develop a simple physical picture for this
massively entangled state, that we dub the Josephson-Kondo screening cloud, and
to propose a realistic setup where its subtle correlations can be unveiled
experimentally.

The use here of the Kondo terminology~\cite{Hewson}, usually associated to the quenching 
of a magnetic moment in a Fermi sea (another example of a complex many-body vacuum) may 
come as a surprise. 
However, deep connections between this purely electronic phenomenon and strongly
dissipative two-level systems have been known for
decades~\cite{Leggett,Weiss,LeHurReview}, a property which we will 
exploit here and further elaborate on. In fact, one challenging and still open question in the
Kondo realm concerns precisely the evidence for the magnetic screening cloud, or in other
words, spatial fingerprints of the entanglement structure between the localized
spin and its surrounding electron 
bath~\cite{Gubernatis,Affleck1,Affleck2,Holzner,Busser,Mitchell}.
Despite numerous experimental efforts~\cite{Bomze,Baines} and innovative theoretical 
proposals~\cite{Affleck3,Cornaglia,Hand,Pereira,Park}, two important difficulties are 
typically encountered. First,
measuring weak long-range magnetic correlations is a daunting task, and
an all-electrical setup would be much preferred. 
%Second, scattering states (in
%2D electron gases, or in 3D bulk metals) are ideally s-wave centered around the
%impurity, but in realistic situations become spatially distorted by the
%electrostatic gates or the underlying disorder, making the precise mapping of the 
%cloud difficult. Engineering the bath as a clean and uniform 1D Fermi liquid
%would certainly be of great advantage, but is hard to achieve with
%standard electron gases. 
%Third, 
Second, spatial Kondo correlations show rapid $2k_F$ Friedel 
oscillations due scattering processes across the Fermi surface, that complicate 
the signal to analyze~\cite{Affleck4}.

\begin{figure}[th]
\includegraphics[width=1.0\linewidth]{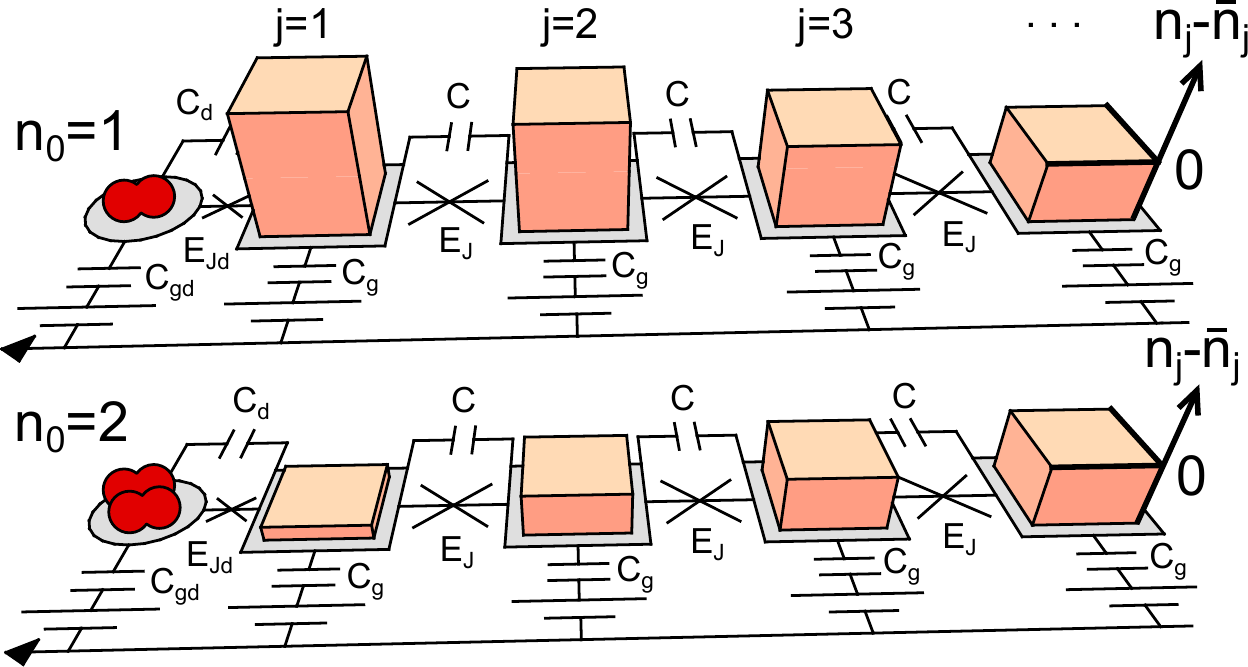}
\caption{(color online) Schematic view of the proposed superconducting circuit
supporting two dressed qubits states, where one or two Cooper pairs in excess on the leftmost 
island are each accompanied by a different charge polarization cloud. Josephson tunneling 
between the two configurations generates entanglement between the qubit and its environment, 
leading to a state that possesses the same singlet structure as the magnetic Kondo screening 
cloud, but in the charge, rather than the spin sector.}
\label{Setup}
\end{figure}
In order to bypass these difficulties, we propose in Fig.~\ref{Setup} a
simple all-superconducting circuit.
Here, one end of a long chain of Josephson elements is coupled to a single superconducting 
quantum dot with a large Coulomb charging energy. The dot is tuned to a charge degeneracy 
point, so that the state space of the dot is reduced to two active charge
levels that contain an excess of either one or two Cooper pairs.
This realizes the analog of a quantized spin 1/2 moment, but in the charge sector, which is
better suited for electrical control with local gates.
%The surrounding environment, in which fermionic quasiparticle excitations are 
%suppressed by the superconducting gap, supports low-energy plasmonic modes which 
%couple both capacitively and inductively (via Josephson tunneling) to the qubit.
The low energy modes of the environment are plasmons that
couple both capacitively and inductively (via Josephson tunneling) to the qubit.
Quenching the charge qubit in this superconducting environment is preferable to quenching
with a normal electron gas because, unlike the capacitive couplings between elements
in the superconducting array, electron-electron interactions in a normal electron gas
act to distort the charge screening cloud. 

The interplay of charging and inductive couplings leads to the following physics. 
The capacitive interaction between the
charge qubit and a high impedance bath generates the dressing of each of the two qubit
states by a distinct charge polarization cloud, as depicted in Fig.~\ref{Setup}. 
The Josephson coupling, rooted in quantum tunneling, forces 
in first approximation a quantum superposition of these two dressed qubit 
states, leading to non-trivial entanglement between the qubit and the bath.
The spatial correlations of this polarization cloud can then be
probed by modulating a comb of local gates applied along the chain, while
recording the charge response of the two-level system. Qubit measurements can be performed  
by weakly coupling the qubit to a superconducting resonator. 
We will show that such measurements allow one to map precisely
the usual magnetic Kondo screening cloud correlations, averaged over fluctuations on the
scale of the Fermi wavelength. This 
bypasses the issue of the $2k_F$ oscillations mentioned above.
A final key aspect pertaining to Kondo physics is universality, namely that phenomena 
beyond a model-specific short distance cut-off are insensitive to microscopic details. 
At low to moderate impedance, we will analytically compute the full spatial
dependence of the Josephson-Kondo screening cloud, and demonstrate its universality
beyond the first few sites of the array.
%at large enough distances.
For large dissipation (circuit impedance roughly larger than half the resistance quantum), we 
will numerically calculate the universal 
%component of the
cloud, using a recently developed technique based on a generic coherent state 
description of environmental wavefunctions, and find that it only depends on the dissipation 
strength and the Kondo length.

{\it Model.}
At temperatures well below the superconducting transition temperature, the proposed Josephson 
circuit is fully governed by the dynamics of conjugate phase $\phi_i$ and charge $n_i$ 
degrees of freedom on the various nodes $i$ of the array, and obeys to the following
Hamiltonian 
\begin{eqnarray}
\nonumber
H&=&\frac{(2e)^2}{2}\sum_{i,j=0}^N (n_i-\bar n_i)\left(\hat C^{-1}\right)_{ij} 
(n_j-\bar n_j)\\
&& +\frac{E_J}{2}\sum_{i=1}^{N-1}(\phi_i-\phi_{i+1})^2  -E_{Jd}\cos(\phi_0-\phi_1),
\label{JJHamiltonian}
\end{eqnarray}
with $\bar n_i$ the average number of Cooper pairs on island $i$, $N+1$ the total number
of islands, and the commutation relation $\left[\phi_j,n_{j'}\right]=i\delta_{jj'}$
(we set $\hbar=1$ in what follows).
The matrix $\hat C$ contains on-site capacitances 
$C_{00}=C_{gd}+C_d$, $C_{11}=C_g+ C_d+C$, $C_{ii}=C_g+2C$, $i\geq 2$,
on the diagonal, as well as nearest neighbor capacitances 
$C_{01}=C_{10}=-C_d$, $C_{i\,i+1}=C_{i+1\,i}=-C$, $i\geq 1$~\cite{Heinzel}. 
All other entries of $\hat C$ are zero.
%The various capacitances are modeled by a matrix~\cite{Heinzel}
%\begin{equation}
%\hat C=\left(\begin{array}{ccccc} 
%C_{gd}+C_d & -C_d & & &\\
%-C_d&C_g+C_d +C & -C & &\\
%& -C & C_g+2C & -C &\\
%& & -C & C_g + 2C &\ddots\\
%& & & \ddots &\ddots
%\end{array}\right).
%\end{equation} 
We assumed that the charging energy in the chain (for sites $i>0$)
is much smaller than the Josephson couplings, namely $(2e)^2/(2 C+C_{g})
\ll E_J$, so that phase differences in the chain are
small. We therefore expanded the Josephson terms $-E_J\cos(\phi_i-\phi_{i+1})$, $i\geq 1$,
to quadratic order. 
We stress that the phase difference between site $i=0$ (the dot), and site $i=1$ of the
chain is not small. This produces the 
anharmonicity in the second line of Eq.~(\ref{JJHamiltonian}), which is
responsible for the non-trivial physics that we discuss now.

A deep connection to quantum optics can be made following three
standard steps~\cite{Goldstein}, starting with a normal mode diagonalization of the quadratic 
part in Eq.~(\ref{JJHamiltonian}), followed by a unitary transformation 
$\widetilde H=UHU^\dagger$, with $U=e^{i (n_0-\bar n_0) \phi_1}$, and
finally a truncation of the Cooper pair box Hilbert space (at site $i=0$) to two
nearly degenerate states, owing to the parametrically large charging energy 
at the end of the chain, namely $(2e)^2/(C_{gd}+C_{d}) \gg E_{Jd}$ 
(see Supplementary Material~\cite{SupInfo} for details). 
Using the replacement with Pauli matrices $(n_0-\bar n_0) \to\sigma_z/2$ and 
$e^{i\phi_0}\to \sigma^+$, and sending $N\to\infty$, this gives readily the Hamiltonian:
\begin{eqnarray}
\label{SBHamiltonian}
\widetilde H & = & \int_0^\pi dk \left[\omega_k b_k^\dagger b_k
-  g_k (b_k^\dagger+b_k)\frac{\sigma_z}{2}\right]
-E_{Jd}\frac{\sigma_x}{2}, \\
\omega_k & = & 2 \sin(k/2)
\sqrt{ \frac{(2e)^2E_J}{C_g+4C\sin^2(k/2)}},\\
g_k & = & 
\frac{1}{\sqrt{2}}\frac{C_{gd}}{C_{gd}+C_d} \frac{\omega_k}{\sin(k/2)} 
\sqrt{\frac{\omega_k}{2\pi E_J}} \cos[k/2-\delta_k],
\end{eqnarray}
the phase shift obeying
$\sin(\delta_k)=\frac{\delta l}{1+\delta l}\sin(\delta_k-k)$, with
$\delta l =C_d C_{gd}/[C_g(C_d+C_{gd})]$. 
Under the assumptions made before on the magnitude of the capacitances, one finds that
$\delta l\ll 1$, so that the phase shift is small in practice.
For low frequencies $\omega \ll \sqrt{(2e)^2E_J/C}$, we find a linear spectral density 
of modes $J(\omega) = 2\pi \alpha \omega$, with effective fine structure
constant (see Supplementary Material~\cite{SupInfo} for a discussion on this 
interpretation of $\alpha$)
\begin{equation}
\label{alpha}
\alpha=\frac{1}{2\pi}\left(\frac{C_{gd}}{C_{gd}+C_d}\right)^2
\sqrt{\frac{(2e)^2}{E_J C}}.
\end{equation}

{\it Protocol.} 
The observable that characterizes the spatial profile of the screening cloud is
\begin{equation}
\chi_j=\left<(n_0-\bar n_0)(n_j-\bar n_j)\right>.\label{clouddef}
\end{equation}  
which quantifies the difference in polarization on site $j$ of the chain, associated with
the distinct charge states $n_0-\bar n_0=\pm 1/2$ of the qubit. 
This quantity can be extracted from frequency-dependent
linear response measurements as follows. We confine ourselves to zero
temperature, as a finite temperature simply introduces exponential decay beyond the 
thermal wavelength~\cite{Borda}.
A small AC gate voltage $V(t)=\delta V e^{i\omega t}$ is applied at the
site $j$ of the chain, causing a perturbation $\delta H = 2e C_g (\hat C^{-1})_{jj}
\delta V e^{i\omega t}(n_j-\bar n_j)$. From the Kubo formula, the linear response of the 
qubit charge at site 0 is
$\big <n_0 -\bar n_0\big>=2 e C_{g}(\hat C^{-1})_{jj} \delta V \chi(j,\omega)$
with $\chi(j,\omega)=-i\int_0^{+\infty} d t e^{i\omega t} 
\big<[n_0(t),n_j(0)]\big>$. At zero temperature if follows from the fluctuation-dissipation
relation that
%, and for $\omega>0$ 
%\begin{equation}
%{\rm Im}\,\chi(j\omega)=-\pi\sum_{n>0}\delta(\omega-E_n+E_0)\left<0\right|n_0\left|n\right>
%\left<n\right|n_j\left|0\right>,
%\end{equation}  
%where $\left|0\right>$ is the many body ground state, $\left|n\right>$ are excited states,
%$E_0$ the ground state energy and $E_n$ the energy of the excited state $\left|n\right>$.
%Integrating over positive frequencies, and using the completeness of the basis 
%$\{\left|n\right>|\,n=0,\,1,\,2,\,\ldots\}$ 
$\chi_j=-\int_0^\infty d\omega\,{\rm Im}\,\chi(j,\omega)/\pi.$

After completing the same transformations on $\chi_j$ used previously to map
Hamiltonian~(\ref{JJHamiltonian}) into (\ref{SBHamiltonian}), one finds from (\ref{clouddef})
the general expression for the Josephson-Kondo cloud:
\begin{eqnarray}
\label{ChiGeneral}
\chi_j &=& \frac{1}{2}\int_0^\pi dk\, 
\theta_{jk}\left<\sigma_z\frac{(b_k^\dagger + b_k)}{\sqrt{2}}\right>
-\frac{\delta_{j1}}{4},\\
\theta_{jk} &=& 4\sqrt{\frac{E_J}{2\pi\omega_k}}\sin(k/2)\cos[k(j-1/2)-\delta_k].
\end{eqnarray}

{\it Regime of intermediate fine structure constant.} For up 
to moderate values of the circuit impedance (typically $\alpha<0.2$), it was
shown recently~\cite{EmeryLuther,Silbey,Harris,Bera1} that the full ground state $\big|\Psi\big>$ describing the coupled 
qubit and Josephson chain system assumes with excellent accuracy a simple 
form $\left|\Psi\right>=\left(\left|f\right>\left|\uparrow\right>+\left|-f\right>\left|\uparrow\right>\right)/\sqrt{2}$, where
\begin{equation}
\label{Silbey}
\left|\pm f\right> =\exp\left[\pm \int_0^\pi dk\, 
f_k(b_k^\dagger-b_k)\right] \left|0\right> ,
\end{equation}
and the oscillators are displaced by:
\begin{equation}
f_k =\frac{1}{2}\frac{g_k}{\omega_k+\Delta_R},~~~
\Delta_R = E_{Jd}\exp\left[-2\int_0^\pi dk\,f_k^2\right].
\label{SilbeyDelta}
\end{equation}
This provides the mathematical foundation for the physical picture shown in
Fig.~\ref{Setup}, underlying the singlet-like structure of the wavefunction,
that leads to entanglement between the qubit and environmental degrees of
freedom.

Substituting the ground state~(\ref{Silbey}) into the Josephson-Kondo cloud
polarizability~(\ref{ChiGeneral}), one finds for $j>1$:
\begin{equation}
\chi_j=- \frac{\Delta_R C_{gd}}{C_{gd}+C_d}
\int_0^\pi \!\! \frac{dk}{2\pi}
\frac{\cos(k/2)\cos[k(j-1/2)-\delta_k]}{\omega_k+\Delta_R}.
\label{exact}
\end{equation}
This expression depends on all microscopic parameters of the device.
However, for $j$ sufficiently larger than $\sqrt{C/C_g}$, 
one finds
%the slowly varying part of the integrand in (\ref{exact}) can be approximated as $1/(E_0k+\Delta_R)$, yielding
universal single-parameter scaling: (See the supplementary
material~\cite{SupInfo} for further detail.)
\begin{equation}
\chi_j=-Z\,{\rm Re}\left\{\,e^{-i\frac{\Delta_R}{E_0}j}
\Gamma\left[0,-i\frac{\Delta_R}{E_0}j\right]\right\},
\label{scaling}
\end{equation}
where $\Gamma(a,z)$ is the incomplete Gamma function, $E_0=\sqrt{(2e)^2 E_J/C_g}$, and
\begin{equation}
Z^{-1}=2\pi\left(1+\frac{C_d}{C_{gd}}\right)\frac{E_0}{\Delta_R}.
\end{equation}
%In formula (\ref{scaling}), the coordinate $j$ should strictly be replaced by  
%$j\to j-1/2+\delta l$ but this small shift is unimportant in the scaling regime, and is therefore
%dropped.

\begin{figure}
\begin{center}
\includegraphics[width=.45\textwidth]{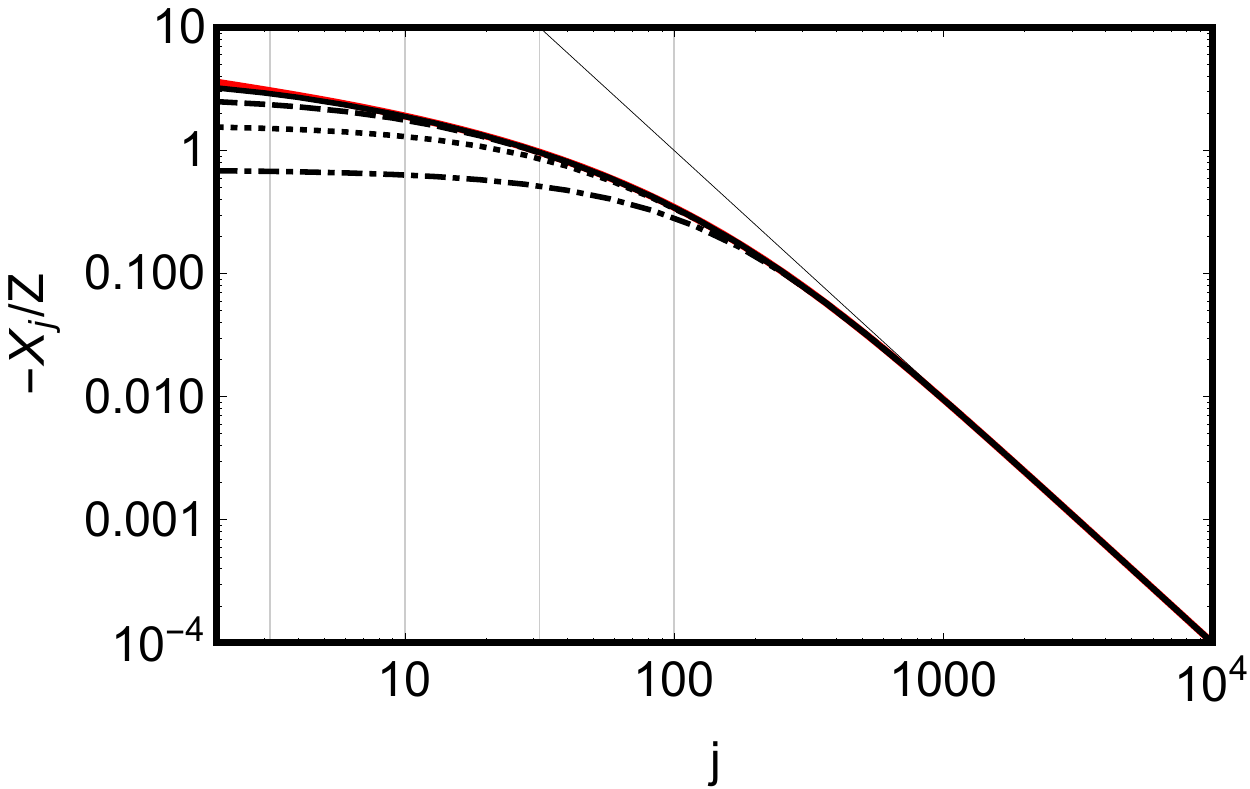}
\caption{Universal Josephson-Kondo cloud at intermediate values of the fine
structure constant $\alpha=0.1$. Here the exact expression~(\ref{exact}) is computed for
several choices of the capacitance ratio in the chain, $C_g/C=10^{-4}$
(dot-dashed), $C_g/C=10^{-3}$ (dotted), $C_g/C=10^{-2}$ (dashed),
$C_g/C=10^{-1}$ (solid), and compared to the analytic scaling
form~(\ref{scaling}). The computation is done for
$E_{Jd}=E_0=0.01$.
Vertical lines at distances $j^\star=\sqrt{C/C_g}=\{\sqrt{10},10,\sqrt{1000},100\}$ 
indicate the cross-over between non-universal short-range and universal long-range 
behavior. 
The thin straight line indicates the long-distance asymptotic behavior 
$-Z \chi_j=(E_0/\Delta_R)^2/j^2$.}
\label{FigScaling}
\end{center}
\end{figure}

The emergence of a universal scaling form can be checked in Fig.~\ref{FigScaling}
comparing the exact expression~(\ref{exact}) to the analytical formula~(\ref{scaling})
for several values of the ratio $C_g/C$ of the chain capacitances.
Formula~(\ref{scaling}) predicts the rapid decay of correlations 
$\chi_j=-Z(E_0/\Delta_R)^2/j^2$ at large distances.   
This result is in full agreement with
the known asymptotics of the magnetic Kondo screening cloud in electronic
systems. This leads to the identification of $E_0/\Delta_R$ as the dimensionless
Kondo length (or Kondo temperature $\Delta_R$). 
The $1/j^2$ behavior may be visible in chains of several thousand sites, that 
can be fabricated with modern lithographic techniques.
%While the absolute signal is weak, Josephson junctions arrays with several
%thousands of sites can be tailored with modern lithographic techniques, making experimental
%investigations possible. 
More interestingly however, the decay at intermediate
distances, in the range $\sqrt{C/C_g} \ll j \ll E_J/\Delta_R$, is much slower, attesting 
to the stronger coupling of the qubit to the Josephson elements in its closer
neighborhood. It is straightforward to see that the existence of these
long-range spatial correlations are intimately connected to the strong entanglement 
between the qubit and its environment. Indeed, a full polarization of the qubit 
into the up state, even dressed by its cloud of oscillators, leads to short
range correlations that do not extend beyond the second site of the chain.
In addition, the long-range correlations of the Josephson-Kondo cloud also reveal 
the non-linearity of the Josephson element coupling site $i=0$ and site $i=1$. 
A purely harmonic chain (see Supplementary Material~\cite{SupInfo}) leads 
to much faster decaying correlations.
The Josephson-Kondo cloud is therefore an experimentally accessible hallmark of
the strong coupling between a two-level system and its macroscopic environment.

\begin{figure}
\begin{center}
\includegraphics[width=.45\textwidth]{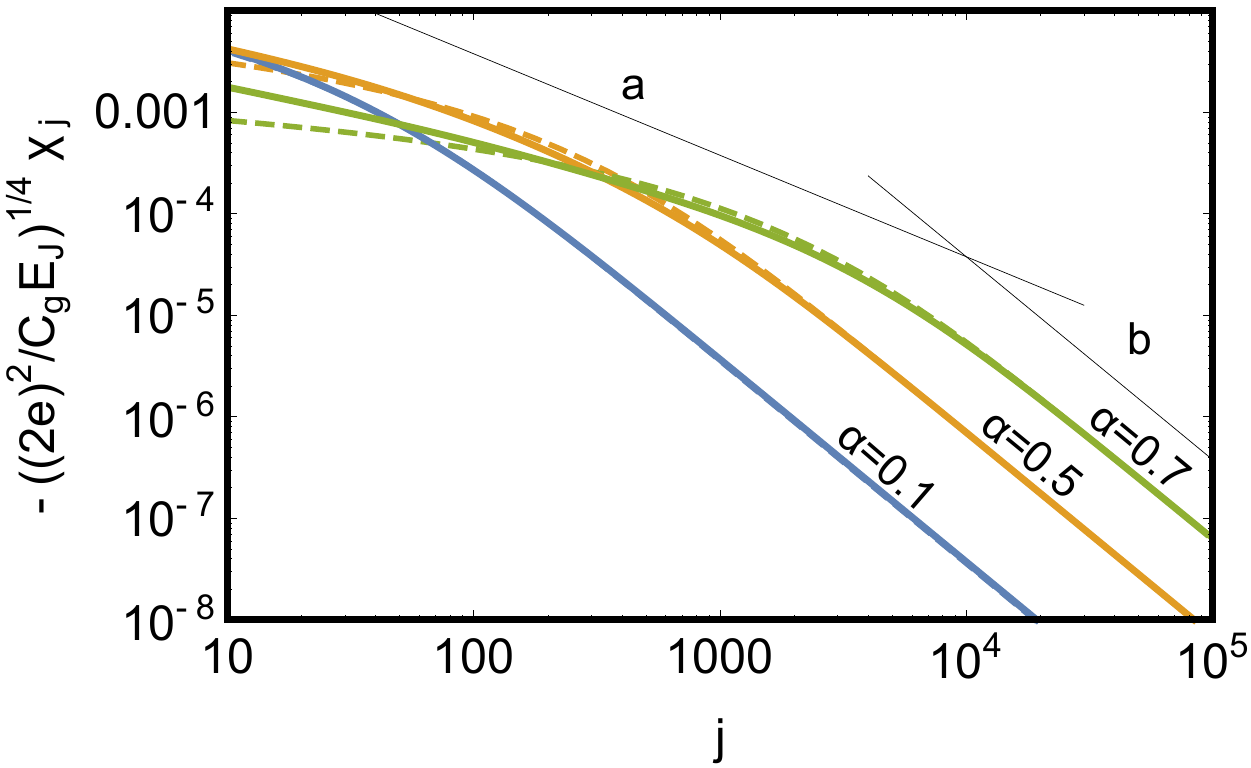}
\caption{Universal Josephson-Kondo cloud at increasing values of the dissipation strength
$\alpha$. Fully converged curves (solid lines) were obtained with $M_\mathrm{cs}=7$ coherent states.
The curves were obtained at the following values of the Kondo energy: 
$\Delta_R=4\times 10^{-2}E_0$ at $\alpha=0.1$, $\Delta_R=4\times 10^{-3}E_0$ at $\alpha=0.5$
and $\Delta_R=5\times 10^{-4}E_0$ at $\alpha=0.7$. In each instance this roughly corresponds to 
$E_{Jd}\sim \sqrt{(2e)^2E_J/C}/20$. The $M_\mathrm{cs}=7$ results are compared
to the $M_\mathrm{cs}=1$ lineshape (\ref{scaling}),
at the same Kondo scale. Lines a $\sim1/j$ and b $\sim 1/j^2$ are there to guide the eye.}
\label{FigLargeAlpha}
\end{center}
\end{figure}

{\it Regime of large fine structure constant.} For circuit impedances close
to the quantum value $h/e^2$, the matter-light interaction parameter $\alpha$
becomes of order one, and the wavefunction~(\ref{Silbey}) is in principle not
sufficient. It was shown recently~\cite{Bera2} that the non-trivial many-body ground state
at large dissipation can be captured by a systematic expansion in terms of coherent states, 
that simply generalizes
the previous ansatz to a superposition of $M_\mathrm{cs}$ coherent states
$\left|\Psi\right> = \sum_{n=1}^{M_\mathrm{cs}} p_n \left|\Psi_n\right>,$
where each $\left|\Psi_n\right>$ is a coherent state of the form (\ref{Silbey}) with
variationally determined amplitudes $p_n$ and displacements $f_k^{(n)}$.
As $M_\mathrm{cs}$ is increased, $\left|\Psi\right>$ rapidly converges to the exact ground state.
Our technique, based on the exact many-body state, provides a powerful alternative to the 
usual numerical renormalization group calculations of the Kondo cloud~\cite{Borda,Lechtenberg}, 
which in contrast, requires a distinct computation for each spatial distance considered.
The resulting correlation function $\chi_j$ for increasing values of the dissipation
strength $\alpha$ is shown in Fig.~\ref{FigLargeAlpha}.
The cloud polarization $\chi_j$ still decays quadratically $\sim - Z (E_0/\Delta_R)/j^2$ at the 
largest distances, but with a Kondo energy $\Delta_R$ that is renormalized downwards from the 
value predicted by (\ref{SilbeyDelta}) (our data at $\alpha=0.7$  shows a
renormalization of $\Delta_R$ by a factor $\sim 1/4$). 
At $j\sim E_0/\Delta_R$, a broad cross-over is visible. In the
intermediate distance universal regime, 
i.e. $\sqrt{C/C_g}\ll j \ll E_0/\Delta_R$, the correlation function $\chi_j$ decays slower 
than $1/j$. At strong dissipation (large $\alpha$), our results are consistent with the perturbative 
prediction $\chi_j\sim-1/[j\ln(j)^2]$, but for the realistic system size
considered, it is difficult to achieve the extreme separation of scales 
between $\sqrt{C/C_g}$ and $E_0/\Delta_R$ that is required to identify this lineshape unambiguously. 
We compare the converged results with $M_\mathrm{cs}=7$ coherent states to the analytical 
$M_\mathrm{cs}=1$ result (\ref{scaling}), 
but with the true Kondo scale $\Delta_R$, rather than the $M_\mathrm{cs}=1$ estimate (\ref{SilbeyDelta}).
It is very interesting that, as the dissipation strength increases, a
noticeable difference develops at intermediate distances between the line shapes predicted 
by the simple $M_{\rm cs}=1$ approximation and the fully converged $M_{\rm
cs}=7$ result.
These differences originate from subtle entanglement within the states in the
bath~\cite{Bera1}.
Indeed, contrary to the simple Ansatz~(\ref{Silbey}) that is based on a single
(multi-mode) coherent state, the environmental wavefunction in the 
multiple coherent state expansion cannot be factorized.

Despite the quantitative differences in the intermediate distance universal regime, it 
is very surprising how accurate the simple Ansatz~(\ref{Silbey}) turns out to
be, highlighting the unexpected simplicity of the many-body ground state of such
a complicated and widely investigated Hamiltonian as the spin-boson model~(\ref{SBHamiltonian}).
Our results indicate that strongly interacting quantum problems display a hidden
structure that is more generic than the specific model studied in this Letter,
since fermionic impurity problems should show similar properties.
With the rapid advances in circuit QED architecture, we believe that future 
experiments following our proposal to probe the Josephson-Kondo cloud will be able
to reveal the simple correlations that are at the heart of \rm{these} many-body states.

Stimulating discussions with Olivier Buisson, Benedikt Lechtenberg and Nicolas Roch 
are gratefully acknowledged. This work is based on research supported in part 
by the National Research Foundation of South Africa (Grant Number 90657).

%% RESET COUNTERS FOR APPENDED SUPPLEMENTARY MATERIAL
\setcounter{figure}{0}
\setcounter{table}{0}
\setcounter{equation}{0}

\onecolumngrid

\global\long\def\theequation{S\arabic{equation}}
\global\long\def\thefigure{S\arabic{figure}}

\vspace{1.0cm}
\begin{center}
{\bf \large Supplementary information for
``Josephson-Kondo screening cloud in circuit quantum electrodynamics''}
\vspace{0.5cm}\\

\begin{quote}
{
\small
\vspace{0.5cm}
We present here for completeness some intermediate steps of the derivation performed in the
main text.
}
\end{quote}
\end{center}

\section{Derivation of the effective spin-boson Hamiltonian}

\subsection{Starting model and definitions}

We start with the Hamiltonian $H=H_0+H_1$, where the quadratic part $H_0$
describes the linearized semi-infinite Josephson junction chain:
\begin{equation}
\label{H0}
H_0=\frac{1}{2}\sum_{m,n=0}^{+\infty} p_m
\left(\hat M^{-1}\right)_{mn}p_n+\frac{1}{2}\sum_{m,n=0}^{+\infty} x_m V_{mn} x_n,
\end{equation}
with $p_m=n_m-\bar n_m$, $x_m=\phi_m$ and $\left[x_m,p_n\right]=i\delta_{mn}$.
The impurity located at the end of the chain contains the only source of
non-linearity of the problem, encoded in the Josephson energy:
$H_1=-E_{Jd}\cos(x_0-x_1)$. The matrix
\begin{equation}
\hat M=\frac{1}{(2e)^2}\left(\begin{array}{cccccc} 
C_{gd}+C_d & -C_d & & & &\\
-C_d&G_g+C_d +C & -C & & &\\
& -C & C_g+2C & -C & &\\
& & -C & C_g + 2C &-C &\\
& & & \ddots & \ddots &\ddots
\end{array}\right)
\end{equation} 
contains the set of capacitances of the whole system, whereas the matrix
\begin{equation}
\hat V=E_J\left(\begin{array}{cccccc} 
0& 0 & & & &\\
0 & 1& -1 & & &\\
&  -1& 2 & -1 & &\\
& & -1 & 2 & -1&\\
& & & \ddots & \ddots &\ddots
\end{array}\right)
\end{equation} 
describes the Josephson couplings in the leads only. The indexing convention is such 
that the upper-left entries of these matrices are labeled by $M_{00}$ and $V_{00}$.
One would like to express $H_0$ as a bath of independent oscillators that are linearly coupled 
to the impurity at site $0$. This is achieved by expressing the set of
harmonic variables in terms of bosonic operators:
\begin{eqnarray}
\label{Defpm}
p_m &=& \frac{1}{\sqrt{2}}\int_0^{+\infty} \!\!\!\!dk\,\, \theta_{mk}(\bk+\bkd) 
\hspace{1cm} m=1,2,\ldots,+\infty\\
\label{Defxm}
x_m &=& \frac{i}{\sqrt{2}}\int_0^{+\infty} \!\!\!\!dk\,\, \xi_{mk}(\bk-\bkd)
\hspace{1cm} m=1,2,\ldots,+\infty
\end{eqnarray}
where $\xi_{mk}$ and $\theta_{mk}$ are real numbers. Note that we adopted the
unusual (but legitimate) convention of ascribing the momentum variable (and
not the position variable) to the sum of creation and annihilation operators.
Since the qubit is capacitively coupled to its environment through the momentum
$p_0$, this choice will result in the standard form of the spin boson
Hamiltonian. We also stress that the coordinates $(x_0,p_0)$ at the end of
the chain are not part of this decomposition, which will faciliate the later
projection onto the two-level system.
A first important relation follows from~(\ref{Defpm}-\ref{Defxm}) by imposing
the commutation relation $[x_m,p_n]=i\delta_{m,n}$ and assuming standard
algebra of the bosonic operators, namely $[\bk,\bkp]=0$,
$[\bk,\bkpd]=\delta(k-k')$:
\begin{equation}
\label{ortho}
\int_0^{+\infty} \!\!\!\!dk\,\, \xi_{mk}\theta_{nk} = \delta_{m,n}.
\end{equation}

\subsection{Normal mode diagonalization of the chain}
Let's focus first on the chain part of $H_0$ which contains the sites 
$m=1,2,\ldots,+\infty$ only, and impose that it is diagonalized by the
bosonic operators:
\begin{equation}
\label{Hchain}
H_0^\mathrm{chain}=\frac{1}{2}\sum_{m,n\neq0} p_m
\left(\hat M^{-1}\right)_{mn}p_n+\frac{1}{2}\sum_{m,n\neq0} x_m V_{mn} x_n
\equiv \int_0^{+\infty} \!\!\!\!dk\,\, \w_k \bkd \bk ,
\end{equation}
with $\w_k$ the positive frequencies of the normal modes. 
Using the decomposition~~(\ref{Defpm}-\ref{Defxm}), this provides several
constraints on the set of unknown parameters:
\begin{eqnarray}
\label{Constraint1}
0 &=& \sum_{m,n\neq0} \left[ \theta_{mk} \left(\hat M^{-1}\right)_{mn} \theta_{nk'} 
-\xi_{mk} V_{mn} \xi_{nl'}\right]\\
\label{Constraint2}
2 \w_k \delta(k-k') &=& \sum_{m,n\neq0} \left[ \theta_{mk} \left(\hat M^{-1}\right)_{mn} \theta_{nk'} 
+\xi_{mk} V_{mn} \xi_{nl'}\right] .
\end{eqnarray}
One can then take the sum and differences of
Eq.~(\ref{Constraint1}-\ref{Constraint2}), and integrate them respectively
with $\int_0^{+\infty} \!\!dk\,\, \xi_{pk'}$ and
with $\int_0^{+\infty} \!\!dk\,\, \theta_{pk'}$. Using the orthogonality
condition~(\ref{ortho}), one readily gets: 
\begin{eqnarray}
\label{EqThetaSum}
\sum_{m\neq0} \theta_{mk} \left(\hat M^{-1}\right)_{mp} &=&
\w_k \xi_{pk} \\
\sum_{m\neq0} \xi_{mk} V_{mp} &=&
\w_k \theta_{pk} .
\end{eqnarray}
In order to write the above equations as full matrix products, we extend
the variables to the range $m=0,1,\ldots,+\infty$, and define the vectors
$\vec \xi_k = (\xi_{0k},\xi_{1k},\xi_{2k},\ldots)$ and
$\vec \theta_k = (\theta_{0k},\theta_{1k},\theta_{2k},\ldots)$.
Noting that $V_{0p}=0$ and with the condition $\theta_{0k}=0$, we can
write:
\begin{eqnarray}
\label{EqTheta}
\hat M^{-1} \vec \theta_k &=& \w_k \vec \xi_{k} \\
\label{EqXi}
\hat V \vec \xi_k &=& \w_k \vec \theta_{k} .
\end{eqnarray}
This results in an eigenvalue equation 
$\hat M^{-1} \hat V\xi_k = \w_k^2 \xi_k$, which can be
rewritten in the explicit form (that avoids inverting explicitly the
matrix $\hat M$):
\begin{equation}
\hat V \vec \xi_k = \w_k^2 \hat M \vec \xi_k .
\end{equation}
This gives a set of conditions, for the different cases $m=0$, $m=1$, and $m\geq2$:
\begin{eqnarray}
\label{xi0}
\xi_{0k} &=& \frac{C_d}{C_{gd}+C_d} \xi_{1k} \\
\label{xi2}
\xi_{2k} &=& \left[ 1- \frac{C_{\mathrm{eff}}\w_k}{(2e)^2E_J-C\w_k^2}\right]\xi_{1k} \\
\label{recursion}
\xi_{m+1,k} &=& 2 \frac{(2e)^2E_J-(C+C_g/2)\w_k^2}{(2e)^2E_J-C\w_k^2} \xi_{mk}
-\xi_{m-1,k}
\end{eqnarray}
with $C_{\rm eff}=C_g+ C_{gd}C_d/(C_{gd}+C_d)$.
Now we exploit the fact that the Josephson chain is uniform except at the first
two sites. We thus introduce scattering states that we parametrize as follows:
\begin{equation}
\label{scattering}
\xi_{mk}=N_k\cos[k(m-1/2)-\delta_k]~~~m=1,\,2,\,\ldots,+\infty
\end{equation}
with amplitude $N_k$ and phase shift $\delta_k$ to be determined. Inserting the
above expression into Eq.~(\ref{recursion}), we readily find the eigenfrequency:
\begin{equation}
\w_k^2=\frac{4(2e)^2E_J\sin^2(k/2)}{C_g+4C\sin^2(k/2)} .
\label{wk}
\end{equation}
The phase shift is then determined from condition~(\ref{xi2}), which reads:
\begin{equation}
\cos\left(\frac{3k}{2}-\delta_k\right) = 
\left[ 1- \frac{C_{\mathrm{eff}}\w_k^2}{(2e)^2E_J-C\w_k^2}\right] 
\cos\left(\frac{k}{2}-\delta_k\right).
\end{equation}
After some trigonometric manipulation and using Eq.~(\ref{wk}), one finds
the simple condition:
\begin{equation}
\sin\left(\delta_k\right) = 
\left[ 1- \frac{C_g}{C_{\mathrm{eff}}}\right] 
\sin\left(\delta_k-k\right),
\end{equation}
which is solved explicitly as:
\begin{equation}
\delta_k = \arctan \left[
\frac{\left(1-\frac{C_g}{C_{\mathrm{eff}}}\right) \sin(k)} 
{\left(1-\frac{C_g}{C_{\mathrm{eff}}}\right) \cos(k)-1} \right].
\end{equation}
It is useful to establish an inverse relation between the harmonic variables
and bosonic operators. Using~(\ref{ortho}), it is straightfoward to check
that the relations~(\ref{Defpm}-\ref{Defxm}) can be inverted by:
\begin{eqnarray}
\bk &=& \frac{1}{\sqrt{2}}\sum_{m=1}^\infty\left[\xi_{mk} p_m - i \theta_{mk} x_m\right] \\
\bkd &=& \frac{1}{\sqrt{2}}\sum_{m=1}^\infty\left[\xi_{mk} p_m + i \theta_{mk} x_m\right] .
\end{eqnarray}
One is equipped now to determine the normalization factor $N_k$, which follows
from the commutation relation $[\bk,\bkpd]=\delta(k-k')$. This relation is equivalent
to the condition
$\vec \theta_k \cdot \vec \xi_{k'}=\delta(k-k')$, which can be rewritten as:
\begin{equation}
\sum_{m,n\neq0} \xi_{mk} V_{mn} \xi_{nk'} = \delta(k-k').
\end{equation}
From the explicit expression of $\hat V$ and the recursion
relations~(\ref{xi2}-\ref{recursion}), one gets:
\begin{equation}
\delta(k-k') = 4 \frac{E_J}{\w_k}\sin^2(k/2)
\left[\frac{C_\mathrm{eff}-C_g}{C_g} \xi_{1k}\xi_{1k'} +
\sum_{m=1}^{+\infty} \xi_{mk} \xi_{mk'} \right].
\end{equation}
Finally, using expression~(\ref{scattering}) and the following standard
algebraic identities
$\sum_{m=1}^{\infty} \cos(mk) = \pi \delta(k) - 1/2$ and 
$\sum_{m=1}^{\infty} \sin(mk) = 1/[2 \tan(k/2)]$, one finds:
\begin{equation}
\delta(k-k') = 2\pi \frac{E_J}{\w_k}\sin^2(k/2) N_k^2 \delta(k-k')+
\mathrm{Finite~Terms},
\end{equation}
where we have singled out the delta-function contribution.
Let us prove that the finite terms are actually zero, which implies the
choice of normalization:
\begin{equation}
N_k=\frac{1}{\sin(k/2)}\sqrt{\frac{\w_k}{2\pi E_J}}.
\end{equation}
For this purpose, let us show that $\vec \theta_k \cdot \vec \xi_{k'} = 0$
for all $k\neq k'$. The proof relies on equations~(\ref{EqTheta}-\ref{EqXi}):
\begin{equation}
\w_{k'}^2 \vec \theta_k \cdot \vec \xi_{k'} = 
\vec \theta_k \cdot \hat M^{-1} \hat{V} \vec \xi_{k'} = 
\left( \hat V \hat M^{-1} \vec\theta_k\right)^{\rm T} \xi_{k'} = 
\w_{k}^2 \vec \theta_k \cdot \vec \xi_{k'}.
\end{equation}
Since $\w_k^2$ is a monotonous function of $k$, the scalar product 
$\vec \theta_k \cdot \vec \xi_{k'}$ indeed vanishes for $k\neq k'$.

\subsection{Expression for the full Hamiltonian}
The impurity part of the quadratic Hamiltonian~(\ref{H0}) related to the site $m=0$ reads:
\begin{equation}
H_0^\mathrm{imp}=\frac{1}{2}\left(\hat M^{-1}\right)_{00}p_0^2
+p_0\sum_{m=1}^{+\infty} \left(\hat M^{-1}\right)_{0m} p_m.
\end{equation} 
Using Eq.~(\ref{Defpm}) and Eq.~(\ref{EqThetaSum}), we finally obtain an exact
and remarkably compact expression for the full Hamiltonian in terms of the normal 
modes and of the variables at the first two sites of the Josephson chain:
\begin{equation}
H= \frac{1}{2}\left(\hat M^{-1}\right)_{00}p_0^2
+\int_0^\pi \!\!\!\!dk\,\, \w_k \bkd \bk
+\frac{p_0}{\sqrt{2}}\int_0^\pi \!\!\!\!dk\,\, \w_k\xi_{0k}(\bk+\bkd)
-E_{Jd}\cos(x_0-x_1).
\end{equation} 
To get rid of the factors containing $x_1$, we perform a unitary transformation 
$\widetilde H=UHU^\dagger$, with $U=e^{ip_0 x_1}$. 
The transformed Hamiltonian reads
\begin{equation}
\widetilde H=\frac{1}{2m}p_0^2 
+\int_0^\pi \!\!\!\!dk\,\, \w_k \bkd \bk
- \frac{1}{\sqrt{2}}p_0\int_0^\pi
\!\!\!\!dk\,\, \w_k\xi_{0k}(\bkd + \bk)-\frac{E_{Jd}}{2}\left[e^{ix_0}+e^{-ix_0}\right]
\end{equation}
where 
\begin{equation}
\frac{1}{m}=\left(\hat M^{-1}\right)_{00}
+\left(1-\frac{2C_d}{C_d+C_{dg}}\right)\int_0^\pi
\!\!\!\!dk\,\, \w_k\xi_{1k}^2.
\end{equation}
Now we assume that we are at a degeneracy point of the uncoupled impurity, and project the impurity part of the Hilbert space onto the two degenerate states.
This results in $p_0\to\sigma_z/2$ and $(e^{ix_0}+e^{-ix_0})\to \sigma_x$. It produces a spin-boson Hamiltonian (where again, we drop constant terms)
\begin{equation}
\tilde H\simeq \int_0^\pi \w_k b_k^\dagger
b_k-\frac{1}{\sqrt{2}}\frac{C_{dg}}{C_{dg}+C_d}\int_0^\pi dk\, \w_k\xi_{1k}(b_k^\dagger+b_k)\frac{\sigma_z}{2}
-E_{Jd}\frac{\sigma_x}{2}
\end{equation}
For $k\ll \sqrt{C_g/C}$, 
\begin{align}
&\w_k\simeq E_0 k\\
&\xi_{1k}\simeq 2\sqrt{\frac{E_0}{2\pi E_J k}}
\end{align}
with $E_0=\sqrt{(2e)^2E_J/C_g}$. We can thus identify the Ohmic spin-boson model parameter
\begin{align}
&\alpha=\frac{1}{2\pi}\left(\frac{C_{dg}}{C_{dg}+C_d}\right)^2\frac{E_0}{E_J}.
\label{Salpha}
\end{align}

From the point of view of the impurity, $\alpha$ is the effective fine structure constant
when the photonic modes of the vacuum are replaced by the plasmonic modes of the superconducting
environment.
This interpretation is confirmed by noting that the fine structure constant 
in vacuum $\alpha_\mathrm{vac.}=(e^2/2h)Z_\mathrm{vac.}$ (with $e$ the electron
charge and $h$ Planck's constant) is fixed to the small 1/137 value by
the vacuum impedance $Z_\mathrm{vac.}= \sqrt{\mu_0/\epsilon_0}\simeq377\, \Omega$. 
In transmission lines, an effective fine structure constant can thus be defined as 
$\alpha=(e^2/2h) Z$ from the environmental impedance $Z$, typically related to a ratio 
from line inductances $L$ and shunt capacitances $C$, $Z=\sqrt{L/C}$. In
superconducting circuits, the Josephson inductance is given by $L=(h/4\pi e)^2 1/E_J$, 
so that we indeed recover Eq.~(\ref{Salpha}) up to numerical and geometrical factors.

\subsection{Connection between the magnetic Josephson-Kondo screening cloud and the 
Kondo screening cloud}
Here we explain the mathematical correspondence between the Kondo screening cloud and the Josephson-Kondo screening cloud.
For a spin $1/2$ magnetic moment at $x=0$, quenched in a one dimensional electron gas, the longitudinal Kondo screening cloud is defined as
\begin{equation}
X^\parallel(x)=4\left<S^{\rm imp}_zS^{\rm el}_z(x)\right>.
\end{equation}
Here $S^{\rm imp}_z$ is the $z$-component of the impurity spin, and $S_z^{\rm el}(x)$ is the $z$-component
of the electron spin density at $x$. 
The cloud can be decomposed into
\begin{equation}
X^\parallel(x)=X_0^\parallel(x)+\cos(2k_Fx) X_{2k_F}^\parallel(x),
\end{equation}
where $X_0^\parallel(x)$ and $X_{2k_F}^\parallel(x)$ vary slowly on the scale of the Fermi wavelength $2\pi/k_F$.
Using the well-known mapping between the Kondo model and the spin-boson model,\cite{Costi,Kotliar} the component $X_0^\parallel(x)$ can be expressed as
\begin{equation}
aX_0^{\parallel}(x)=\frac{1}{\pi}\int_0^\infty dq\,X_q(x)\left<\sigma_z\frac{(b_q^\dagger + b_q)}{\sqrt{2}}\right>\label{xk}
\end{equation}
where 
\begin{equation}
X_q(x)=\sqrt{q}e^{-q/2}\cos\left(\frac{qx}{a}\right).
\end{equation}
Here $1/a$ is the ultra-violet scale in the problem (of the order of $k_F$), $q$ is dimensionless,  and
the expectation value is with respect to the ground state of the spin-boson model. 
Comparing this expression to the expression (7) for $\chi_j$ in the main text, we see that the two integrands
have the same low $q$ behavior. Thus, the $0k_F$ component of the Kondo cloud for $x\gg a$, 
corresponds to the Josephson-Kondo cloud for $j\gg\sqrt{C/C_g}$. Explicitly, the correspondence is
\begin{align}
& aX_0^\parallel(x)\leftrightarrow\sqrt{\frac{2E_0}{\pi E_J}}\chi_j\nonumber\\
&\frac{x}{a}\leftrightarrow j-\frac{1}{2}+\delta l\label{map}
\end{align}

At distances where the correspondence is accurate, the profile of the cloud is universal. This can be seen analytically in the regime of small to moderate dissipation, where the exact Josephson-Kondo cloud is
given by Eq.~(11) in the main text. As pointed out there, it depends on all the microscopic parameters of
the device. However, for $j$ sufficiently larger than $\sqrt{C/C_g}$, but not necessarily larger than 
$1/\Delta_R$, the integrand of Eq.~(11) in the main text can be approximated as
\begin{equation}
\frac{\cos[k(j-1/2+\delta l)]}{E_0 k+\Delta_R},
\end{equation}  
due to rapid oscillations when $k\gtrsim 1/j$. For the same reason, the upper bound of the integral can be extended from $\pi$ to $\infty$. Under this approximation, the integral evaluates to
\begin{equation}
\chi_j=-Z{\rm Re}\left\{e^{-i\frac{\Delta_R}{E_0}(j-1/2+\delta l)}\Gamma\left[0,-i\frac{\Delta_R}{E_0}(j-1/2+\delta l)\right]\right\},
\end{equation}
where $\Gamma(a,z)$ is the incomplete Gamma function and $Z$ is as defined in Eq.~(13) in the main text.
We have said that under the assumptions we make about capacitances, $\delta l$ is small. In the same 
regime, the Kondo scale $\Delta_R$ is small relative to the ultra-violet scale $E_0$. This implies
that $\chi_j$ varies slowly on the scale of $\delta_l-1/2$, and we can therefore replace 
$j-1/2+\delta l\to j$ in the above equation. In this way, we obtain Eq.~(12) in the main text. 

\subsection{The harmonic chain}
In the main text we argued that the non-trivial physics we investigate is generated by the
anharmonic term $-E_{Jd}\cos(\phi_0-\phi_1)$ in the Hamiltonian (1).
Specifically, we want to show that without the anharmonicity, correlations decay 
far more rapidly than with the anharmonicity.
In order to justify these statements and make them more precise, we study here the harmonic chain.
That is, we consider the regime $E_{Jd}\gg(2e)^2/(C_d+C_{gd})$, where the cosine term in the Hamiltonian (1) can be expanded to quadratic order in the phase difference $\phi_0-\phi_1$, yielding a
harmonic chain 
\begin{equation}
H_H=\frac{(2e)^2}{2}\sum_{i,j=0}^\infty (n_i-\bar n_i)\left(\hat C^{-1}\right)_{ij} 
(n_j-\bar n_j)+\frac{E_J}{2}\sum_{i=1}^{\infty}(\phi_i-\phi_{i+1})^2  +\frac{E_{Jd}}{2}
(\phi_0-\phi_1)^2.
\label{HHamiltonian}
\end{equation}     
The Hamiltonian is again diagonalized by defining bosonic operators $b_k$ and 
$b_k^\dagger$, for $k\in[0,\pi]$,
such that 
\begin{equation}
b_k=\frac{1}{\sqrt{2}}\sum_{j=0}^\infty\left[\xi_{jk}(n_j-\bar n_j)-i\theta_{jk}\phi_j\right].
\end{equation}
As before, $\xi_{jk}$ is obtained from the classical equations of motion. The presence of the quadratic
Josephson coupling between sites $i=0$ and $i=1$ modifies the expressions for $\xi_{jk}$ as follows.
The relation between $\xi_{0k}$ and $\xi_{1k}$ now reads
\begin{equation}
\xi_{0k}=(1-F_k)\xi_{1k},~~~F_k=\frac{\omega_k^2 C_{gd}}{\omega_k^2(C_{gd}+C_d)-(2e)^2E_{Jd}}.
\end{equation}
For $j\geq 1$, the form of $\xi_{jk}$ is still
\begin{equation}
\xi_{jk}=N_k\cos[k(j-1/2)-\delta_k].
\end{equation}
In the last two equations, the from of $\omega_k$ and $N_k$ are unchanged from what they were before, but
the phase shift $\delta_k$ is obviously affected by the Josephson coupling between sites $i=0$ and
$i=1$. It is now given by
\begin{equation}
\tan(\delta_k)=\frac{\sin k}{\cos k-1-\frac{C_g}{C_{gd}(1-F_k)}}.
\end{equation}
The relationship between $\vec{\theta}_k$ and $\vec{\xi}_k$ is still given by
(S12) and (S13), but in (S13), one has to remember to include the extra $E_{Jd}$ couplings, i.e.
 $V_{00}=E_{Jd}$, $V_{11}=E_{Jd}+E_{J}$, and $V_{01}=V_{10}=-E_{Jd}$.
The ground state of the harmonic chain (\ref{HHamiltonian}) is the bosonic
vacuum $\left|0\right>$ defined by the condition that
$b_k\left|0\right>=0$ for all $k$ in $[0,\pi]$.
 
 \begin{figure}[thb]
\begin{center}
\includegraphics[width=.7\textwidth]{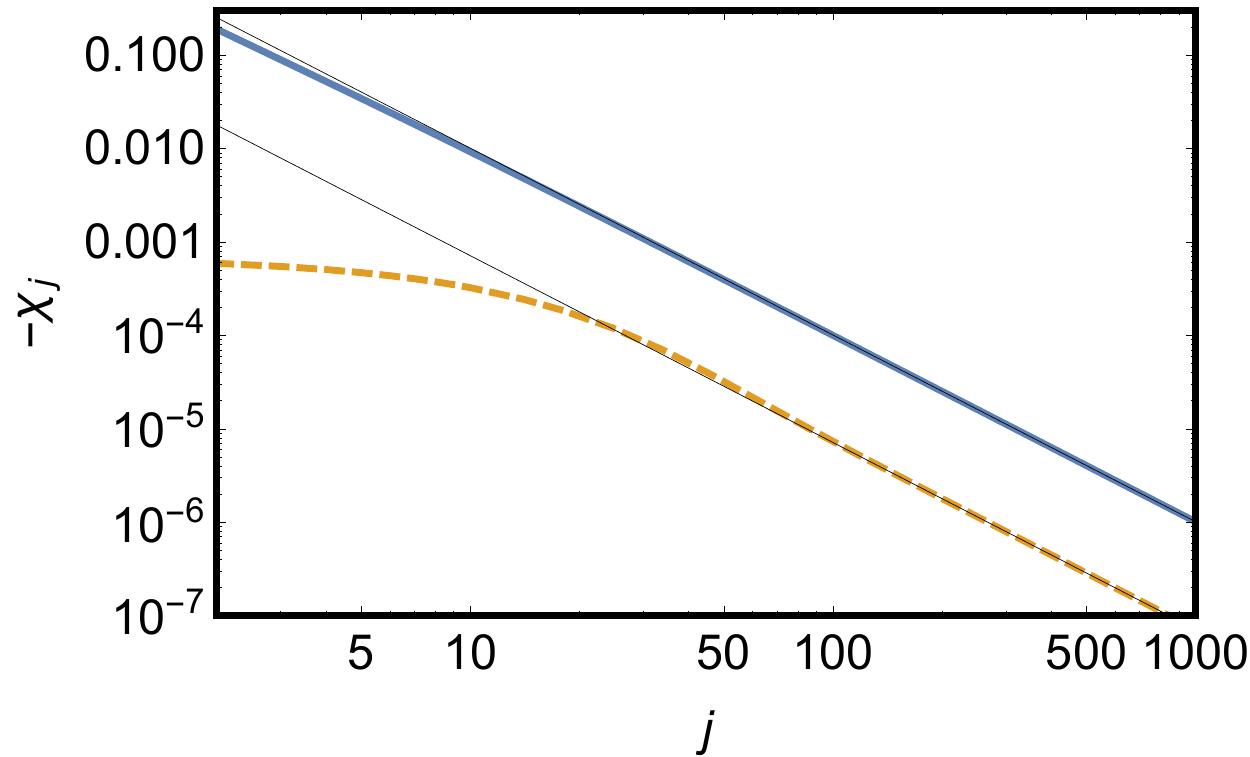}
\caption{The correlation function $\chi_j$ for the harmonic chain,  calculated from Formula (\ref{HarmonicChi}). The dashed curve shows the result for weak dissipation (large $C$), and 
was calculated for $C_d=C_{gd}=C_g$, $E_{Jd}=10 (2e)^2/C_g$, and $E_J=10 (2e)^2/(2C+C_g)$.
For the anharmonic impurity system (i.e. a much smaller value of $E_{Jd}$), the remaining parameter
values would have corresponded to $\alpha=0.035$. The solid curve shows the result for strong dissipation (small $C$), and 
was calculated for $C=C_d=C_g/100$, $C_{gd}=C_g$, and $E_{Jd}=E_{J}=10 (2e)^2/C_g$.
For the anharmonic impurity system (i.e. a much smaller value of $E_{Jd}$), in this case, 
the remaining parameter
values would have corresponded to $\alpha=0.5$. For both the solid and the dashed curve, a thin solid
line indicates $1/j^2$ with the prefactor predicted by Formula (\ref{approx2}).  
\label{fs1}}
\end{center}
\end{figure}

For the correlator $\chi_j$ one then obtains
\begin{eqnarray}
\chi_j&=&\left<0\right|(n_0-\bar n_0)(n_j-\bar n_j)\left|0\right>\nonumber\\
&=&\frac{1}{2}\int_0^\pi dk\,\theta_{0k}\theta_{jk}\nonumber\\
&=&-\frac{E_J}{\pi}\int_0^\pi dk\,\frac{F_k}{\omega_k}\cos(k/2-\delta_k)\cos[k(j-1/2)-\delta_k],
\label{HarmonicChi}
\end{eqnarray}
the last line being valid for $j\geq 2$. Note that, while $F_k$ diverges at the point $k=k_*$ such that
$\omega_{k_*}^2=(2e)^2E_{Jd}/(C_{gd}+C_d)$, the product $F_k\cos(k/2-\delta_{k})$ remains finite
at $k=k_*$, so that the integrand in the last line of (\ref{HarmonicChi}) is a smooth function of $k$.
At sufficiently large $j$, the slowly varying factor $F_k\cos(k/2-\delta_k)/\omega_k$ can be evaluated
to linear order in $k$, yielding
\begin{eqnarray}
\chi_j&\simeq&\frac{1}{\pi}\frac{E_0C_{gd}}{(2e)^2}\int_0^\pi dk\,k\cos[k(j-1/2)-\delta_k],\nonumber\\
&=&\frac{1}{\pi}\frac{E_0C_{gd}}{(2e)^2}\partial_j\int_0^\pi dk\,\sin[k(j-1/2)-\delta_k],
\label{approx1}
\end{eqnarray}
At large $j$, the phase shift $\delta_k$ dephases the contribution from evaluating the integral without the phase shift at $k=\pi$. 
Replacing $j-1/2\to j$, which is allowed within the accuracy to which we evaluated the integral, we
then find
\begin{eqnarray}
\chi_j&\simeq&\frac{1}{\pi}\frac{E_0C_{gd}}{(2e)^2}\partial_j
\left[-\frac{1}{j}\cos(jk)\right]_{0}^{\pi}=-\frac{E_0C_{gd}}{\pi(2e)^2}\frac{1}{j^2}.
\label{approx2}
\end{eqnarray}
For the anharmonic system that we studied in the main text, we found on the other hand
\begin{equation}
\chi_j\simeq-Z\left(\frac{E_0}{\Delta_R}\right)^2\frac{1}{j^2}=- \frac{(2e)^2}{2\Delta_R(C_d+C_{gd})}\times\frac{E_0C_{gd}}{\pi(2e)^2}
\frac{1}{j^2}.
\end{equation}
At strong dissipation, this results in a correlation function that is larger than 
in the harmonic case by a factor proportional to the inverse of the (exponentially small) Kondo 
energy.

Next we ask at what distance the $1/j^2$ behavior predicted in Formula (\ref{approx2}) sets in.  For the harmonic chain, the $1/j^2$ decay breaks down below 
a distance $j\sim1/k_0$, where $k_0$ is the value of $k$ where the $k^3$ term in the Taylor expansion
of $F_k\cos(k/2-\delta_k)/\omega_k$ in the integrand of (\ref{HarmonicChi}) becomes important.
Under the assumptions we make about the sizes of capacitances, it follows that $k_0\sim\min\{\sqrt{C_g/C},1\}$.
Hence, for the harmonic chain, $\chi_j\sim 1/j^2$ behavior is seen for $j\gg\max\{\sqrt{C/C_g},1\}$. 
In contrast, for the anharmonic impurity, we saw in the main text that there is an intermediate regime
$\sqrt{C/C_g}\ll j \ll E_0/\Delta_R$, where correlations decay far more slowly.

We also note the following. In the anharmonic impurity problem, one increases the dissipation strength
$\alpha$ to values of order $1$ by decreasing $C$, while keeping $E_J(2C+C_g)/(2e)^2$ large. This has
the effect of decreasing $\Delta_R$, and hence the region of the onset of $1/j^2$ decay is shifted to
larger $j$. In the harmonic problem, increasing $C$ tends to have the opposite effect, namely, it shifts
the onset of $1/j^2$ decay to smaller $j$. This is illustrated in Fig.~\ref{fs1}, where the exact 
formula (\ref{HarmonicChi}) and the large $j$ asymptotic formula (\ref{approx2}) for $\chi_j$ in the 
harmonic system are compared, for
fixed $E_J(2C+C_g)/(2e)^2$, but different $C$.  
These results
justify the statement we make in the main text that the Josephson-Kondo cloud is a hallmark of
the strong coupling between a two-level system and its macroscopic environment.  


\begin{thebibliography}{99}
\frenchspacing

% Review on superconducting qubits
\bibitem{Makhlin} Y. Makhlin, G. Sch\"on, and A. Shnirman,
Rev. Mod. Phys. {\bf 73}, 357 (2001).

% Circuit QED Review
\bibitem{Schoelkopf} R. J. Schoelkopf and S. M. Girvin,
Nature {\bf 451}, 664 (2008).

% Josephson metamaterials
\bibitem{Zueco} D. Zueco, C. Fern\'andez-Juez, J. Yago, U. Naether, B. Peropadre, J. J.
Garc\'ia-Ripoll, and J. J. Mazo, Supercond. Sci. Technol. {\bf 26} 074006 (2013).
\bibitem{Jung} P. Jung, A. V. Ustinov, S. M. Anlage,
Supercond. Sci. Technol. {\bf 27}, 073001 (2014).

% Various recent experiments in circuit QED
\bibitem{Niemczyk} T. Niemczyk, F. Deppe, H. Huebl, E. P. Menzel, F. Hocke, 
M. J. Schwarz, J. J.  Garc\'ia-Ripoll, D. Zueco, T. H\"ummer, E. Solano, A. Marx, 
and R. Gross, Nat. Phys. {\bf 6}, 772 (2010).
\bibitem{Forn} P. Forn-D\'iaz, J. Lisenfeld, D. Marcos, J. Garc\'ia-Ripoll, E. Solano, 
C. Harmans, and J. Mooij, Phys. Rev. Lett. {\bf 105}, 237001 (2010).
\bibitem{Astafiev} O. Astafiev, A. M. Zagoskin, A. A. Abdumalikov, Y. A.
Pashkin, T. Yamamoto, K. Inomata, Y. Nakamura, and J. S. Tsai, Science {\bf 327}, 
840 (2010).
\bibitem{Abdumalikov} A. A. Abdumalikov, O. Astafiev, A. M. Zagoskin, Y. A. Pashkin, Y.
Nakamura, and J. S. Tsai, Phys. Rev. Lett. {\bf 104}, 193601 (2010).
\bibitem{Sanchez} E. S\'anchez-Burillo, D. Zueco, J. Garc\'ia-Ripoll,
L. Martín-Moreno, Phys. Rev. Lett. {\bf 113}, 263604 (2014).


%Kondo resonance of a microwave photon 
\bibitem{LeHur} K. Le Hur, {Phys. Rev. B} {\bf 85}, 140506 (2012).
 
%Inelastic microwave photon scattering off a quantum impurity in a Josephson-junction array
\bibitem{Goldstein} M. Goldstein, M. H. Devoret, M. Houzet, and L. I. Glazman, 
{Phys. Rev. Lett.} {\bf 110}, 017002 (2013).

% Multipolaron Ansatz
\bibitem{Bera1} S. Bera, S. Florens, H. U. Baranger, N. Roch, A. Nazir,
and A. W. Chin, Phys. Rev. B {\bf 89}, 121108(R) (2014).

% Kondo book
\bibitem{Hewson} A. C. Hewson, {\it The Kondo Problem to Heavy Fermions} (Cambridge
University Press, Cambridge, UK, 1993).

% Reviews on dissipative two-state system
\bibitem{Leggett} A. J. Leggett, S. Chakravarty, A. T. Dorsey, M. P. A. Fisher, 
A. Garg, and W. Zwerger, {Rev. Mod. Phys.} {\bf 59}, 1 (1987).
\bibitem{Weiss} U. Weiss, {\it Quantum Dissipative Systems} (World Scientific, 1993).
\bibitem{LeHurReview} K. Le Hur, {Ann. Phys.} {\bf 323}, 2208 (2008).

% Experiment on Kondo box
\bibitem{Bomze} Yu. Bomze, I. Borzenets, H. Mebrahtu, A. Makarovski, H. U.
Baranger, and G. Finkelstein, Phys. Rev. B {\bf 82}, 161411 (2010).
\bibitem{Baines} D. Y. Baines, T. Meunier, D. Mailly, A. D. Wieck, C. B\"auerle,
L. Saminadayar, P. S. Cornaglia, G. Usaj, C. A. Balseiro, and D. Feinberg, Phys.
Rev. B {\bf 85}, 195117 (2012).

% Kondo cloud basics
\bibitem{Gubernatis} J. E. Gubernatis, J. E. Hirsch, and 
D. J. Scalapino, Phys. Rev. B {\bf 35}, 8478 (1987).
\bibitem{Affleck1} V. Barzykin and I. Affleck,
Phys. Rev. Lett. {\bf 76}, 4959 (1996).
\bibitem{Affleck2} I. Affleck, in {\em Perspectives of Mesoscopic Physics: Dedicated to Yoseph Imry's 70th Birthday}, A. Aharony and O. Entin-Wohlman (eds.) (World Scientific, Singapore, 2010).
\bibitem{Holzner} A. Holzner, I. P. McCulloch, U. Schollw\"ock, 
J. von Delft, and F. Heidrich-Meisner, Phys. Rev. B {\bf 80}, 205114 (2009).
\bibitem{Busser} C. A. B\"usser, G. B. Martins, L. Costa Ribeiro, E. Vernek, 
E. V.  Anda, and E. Dagotto, Phys. Rev. B {\bf 81}, 045111 (2010).
\bibitem{Mitchell} A. K. Mitchell, M. Becker, and R. Bulla, Phys. Rev. B 84,
115120 (2011).

% Kondo cloud altenative proposals
\bibitem{Affleck3} I. Affleck and P. Simon,
Phys. Rev. Lett. {\bf 86}, 2854 (2001).
\bibitem{Cornaglia} P. S. Cornaglia and C. A. Balseiro,
Phys. Rev. Lett. {\bf 90}, 216801 (2003).
\bibitem{Hand} T. Hand, J. Kroha, and H. Monien,
Phys. Rev. Lett. {\bf 97}, 136604 (2006).
\bibitem{Pereira} R. G. Pereira, N. Laflorencie, I. Affleck, and B. I. Halperin,
Phys. Rev. B {\bf 77}, 125327 (2008).
\bibitem{Park} J. Park, S.-S. B. Lee, Y. Oreg, and H.-S. Sim,
Phys. Rev. Lett. {\bf 110}, 246603 (2013).
\bibitem{Bauer} J. Bauer, C. Salomon, and E. Demler,
Phys. Rev. Lett. {\bf 111}, 215304 (2013).

% Kondo cloud: Friedel oscillations
\bibitem{Affleck4} I. Affleck, L. Borda, and H. Saleur, Phys. 
Rev. B {\bf 78}, 139902 (2008). 

% Capacitance matrix
\bibitem{Heinzel} T. Heinzel, {\it Mesoscopic Electronics in
Solid State Nanostructures} (Wiley, Weinheim, Germany, 2007).

% Variational calculations for spin-boson model
\bibitem{EmeryLuther} V. J. Emery and A. Luther, {Phys. Rev. Lett.} {\bf
26}, 1547 (1971).
\bibitem{Silbey} R. Silbey and R. A. Harris, 
{J. Chem. Phys.} {\bf 80}, 2615 (1984); 
\bibitem{Harris} R. A. Harris and R. Silbey, J. Chem. Phys. {\bf 83}, 1069 (1985).

% Kondo cloud: finite T
\bibitem{Borda} L. Borda, Phys. Rev. B {\bf 75}, 041307(R) (2007).

% General multipolaron expansion
\bibitem{Bera2} S. Bera, A. Nazir, A. W. Chin, H. U. Baranger, 
and S. Florens, Phys. Rev. B {\bf 90}, 075110 (2014).

% Kondo cloud dynamics
\bibitem{Lechtenberg} 
B. Lechtenberg and F. B. Anders, Phys. Rev. B {\bf 90}, 045117 (2014).

% Supplementary Material
\bibitem{SupInfo} Supplementary Material document.

\end{thebibliography}

\begin{thebibliography}{99}
\frenchspacing

%Mapping between Kondo and Spin-boson models
\bibitem{Costi} T. A. Costi and G. Zar\'and, Phys. Rev. B {\bf 59}, 12398, (1999).
\bibitem{Kotliar} G. Kotliar and Q. Si, Phys. Rev. B {\bf 53}, 12373, (1996).
\end{thebibliography}
\end{document}